\documentclass[ twocolumn, superscriptaddress, prx]{revtex4-1}

\usepackage{graphicx}
\usepackage{amsmath,amssymb}

\begin{document}


\title{Molecular behavior of DNA in a cell-sized compartment coated by lipids}

\author{Tsutomu Hamada}
 \email{t-hamada@jaist.ac.jp}
\author{Rie Fujimoto}
\affiliation{
School of Materials Science, Japan Advanced Institute of Science and Technology, 1-1 Asahidai, Nomi, Ishikawa 923-1292, Japan
}

\author{Shunsuke F. Shimobayashi}
\author{Masatoshi Ichikawa}
\affiliation{
Department of Physics, Graduate School of Science, Kyoto University, Kyoto 606-8502, Japan
}

\author{Masahiro Takagi}
 \email{takagi@jaist.ac.jp}
\affiliation{
School of Materials Science, Japan Advanced Institute of Science and Technology, 1-1 Asahidai, Nomi, Ishikawa 923-1292, Japan
}

\date{\today}
\date{\today}
\begin{abstract}
The behavior of long DNA molecules in a cell-sized confined space was investigated. We prepared water-in-oil droplets covered by phospholipids, which mimic the inner space of a cell, following the encapsulation of DNA molecules with unfolded coil and folded globule conformations. Microscopic observation revealed that the adsorption of coiled DNA onto the membrane surface depended on the size of the vesicular space. Globular DNA showed a cell-size-dependent unfolding transition after adsorption on the membrane. Furthermore, when DNA interacted with a two-phase membrane surface, DNA selectively adsorbed on the membrane phase, such as an ordered or disordered phase, depending on its conformation. We discuss the mechanism of these trends by considering the free energy of DNA together with a polyamine in the solution. The free energy of our model was consistent with the present experimental data. The cooperative interaction of DNA and polyamines with a membrane surface leads to the size-dependent behavior of molecular systems in a small space. These findings may contribute to a better understanding of the physical mechanism of molecular events and reactions inside a cell.
\end{abstract}

\maketitle

The principles that govern molecular systems to generate living organisms are still an unsolved problem in soft matter and/or biological physics. Cells are enclosed by a soft membrane interface with lipid bilayers \cite{Cell}. The lipid membranes produce a micrometer-scale confined space to encapsulate biological macromolecules, such as genomic DNA and proteins. Within such a small space, the membrane surface has a significant effect on the inner molecular dynamics because of an increase in the surface/volume ratio. To understand the characteristics of biological molecular systems that function within a cell, studies under conditions of a cell-sized compartment are invaluable. Recently, cell-sized lipid vesicles have been used in studies to reveal macromolecular behaviors inside a cell \cite{Materials2012,Fletcher}, which leads to research in related areas, such as on synthetic materials that mimic biological cells \cite{Science2014}. Several molecular systems have been found to change their functions or structural forms within a cell-sized compartment covered by membranes \cite{Nomura2003, Takiguchi2011, Negishi2011, Shew}.



DNA is a representative of biological macromolecules that are present in a cell. The organization and structural dynamics of DNA are essential for the regulation of active cellular systems, such as protein synthesis and cell division \cite{Cell}. A DNA molecule can be characterized as a negatively-charged semiflexible polymer. A long DNA molecule that is larger than several tens of kilo basepairs (kbp) shows a discrete conformational transition between an unfolded coil state and a folded globule state, depending on the presence of condensing agents such as polyamines; the molecular behavior can be described by free energy including electrostatic interactions, the conformational entropy of DNA and the translational entropy of counterions \cite{Baigl2011,b,Bloomfield}. The difference between these two conformations is important for biological functions, such as gene expression \cite{Yamada2005}. It is also known that DNA molecules partly localize near the nuclear membrane surface in a cell \cite{Marshall1996}. Recently, using a cell-sized lipid vesicle, Kato et al. reported that DNA inside the vesicle changes its conformation and shows membrane-adsorption in the presence of Mg$^{2+}$ ion \cite{Kato2009}. However, the physical mechanisms that underlie the molecular organization of DNA in a cell space are still poorly understood.

\begin{figure}[htbp]
\centering
\includegraphics{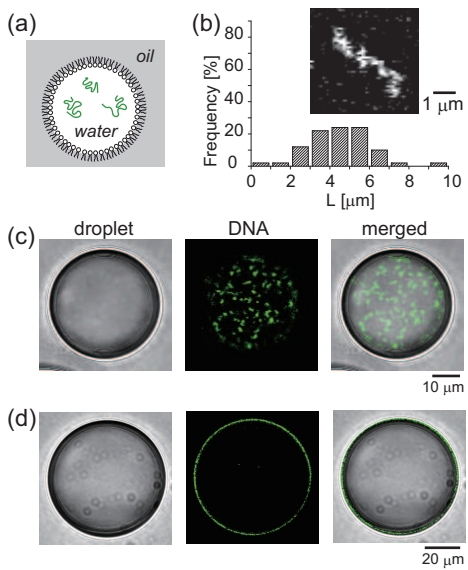}
\caption{\label{1} (a) Schematic illustrations of a cell-sized droplet that encapsulates DNAs. (b) Long-axis length ($L$) of unfolded coiled T4 DNA together with a typical microscopic image. The solution was 10 mM Tris-HCl. N=50.  (c, d) Behavior of coiled DNA with SPD at $\leq$0.5 mM in droplets with DOPC membrane. DNA diffused in an aqueous phase in a small droplet (c), whereas DNA adsorbed on the surface in a large droplet (d). SPD concentrations in c and d were the same (0.2 mM).}
\end{figure}

We performed the microscopic observation of Bacteriophage T4 DNA encapsulated in a cell-sized droplet with a change in the concentration of a condensing agent, such as a trivalent positively charged ion, spermidine (SPD)  (Fig. 1(a)) \cite{microscope, JPCL2010}.  
We first focus on the molecular behavior of coiled DNA under low concentrations of SPD ($<$ 1 mM). Figure 1(b) shows a typical microscopic image of a single molecule of coiled DNA, and the distribution of the long-axis length $L$. Without SPD, coiled DNA molecules did not adhere to the membrane surface, and showed Brownian motion in the inner aqueous phase. In the presence of 0.2 or 0.5 mM SPD, DNA was adsorbed onto the membrane surface (Fig. 1(d)). We found that the preference for DNA adsorption depends on the size of the droplets. In small droplets with a radius below around 30 $\mu$m, DNA did not adsorb on, but rather diffused inside, the droplets (Fig. 1(c)), while DNA tended to adsorb on the surface of large droplets (Fig. 1(d)) \cite{size}. 
\begin{figure}[htbp]
\centering
\includegraphics{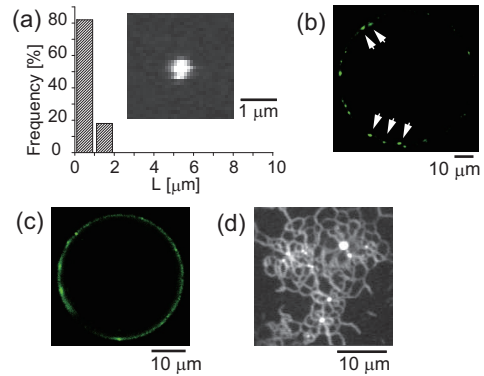}
\caption{\label{1} (a) Long-axis length ($L$) of folded globular T4 DNA together with a typical microscopic image. The solutions were 10 mM Tris-HCl with 2.5 mM SPD. N=50. (b-d) Behavior of globular DNA with SPD at $\geq$1.0 mM in droplets with a DOPC membrane. Typical images of DNA before (b) and after (c) incubation for 2 h, using a laser scanning microscope. (d) An aggregation structure of DNA on the membrane surface after 2 h of incubation, which was observed with a fluorescent microscope. SPD concentrations in c and d were the same (2.5 mM).}
\end{figure}

Above 1.0 mM SPD, DNA essentially shows a globular conformation in bulk (Fig. 2(a)). When the globular DNA solution was encapsulated in the droplets, DNA first adsorbed on the membrane surface while retaining a globular conformation (Fig. 2(b)). Several minutes after adsorption, the adsorbed DNA showed an unfolding transition from a globule to a coil (Fig. 2(c)). This unfolding behavior appeared in droplets with a radius larger than approximately 30 $\mu$m, while DNAs in smaller droplets retained their globular shape. Subsequently, a network or a web-shaped aggregation of DNA molecules on the membrane surface was observed in large droplets (Fig. 2(d)). 


\begin{figure}[htbp]
\centering
\includegraphics{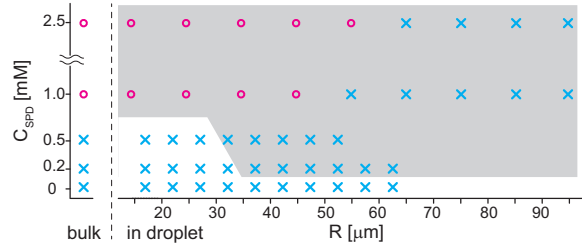}
\caption{\label{1} Phase diagram of DNA behavior in droplets as a function of droplet radius $R$ and SPD concentration $C_\text{SPD}$. Crosses and circles indicate coil and globular conformations of DNA, respectively. Gray region corresponds to the adsorption of DNA on the membrane surface.}
\end{figure}

A quantitative phase diagram is presented in Fig. 3 \cite{note2}. In bulk, DNA exhibits a conformational transition from a coil to a globule as the SPD concentration exceeds 1.0 mM. When coiled DNAs were encapsulated within droplets, DNA tended to adsorb on the surface of droplets with a radius above around 30 $\mu$m. DNA in the 0.5 mM SPD solution was adsorbed on the droplet surface more frequently than that within a 0.2 mM SPD solution. Thus, SPD increased the frequency of DNA adsorption. When globular DNAs were encapsulated, DNAs adsorbed on the surface and exhibited unfolded behavior, including $\prime$globule to coil$\prime$ and $\prime$globule to web$\prime$ transitions. In addition, the unfolding of DNA on membranes was shown to depend on the size of the droplets. DNA that adhered to the inner surface of large droplets tended to exhibit an unfolding transition, whereas DNA remained in the globule state on the surface of small droplets. Coiled DNA, which was unfolded from a globule on the membrane surface, often exhibited a network or web pattern (Fig. 2(d)), which is different from the adsorption of coiled DNA with lower concentrations of SPD (0.2 or 0.5 mM). 



\begin{figure}[htbp]
\centering
\includegraphics[scale=0.35]{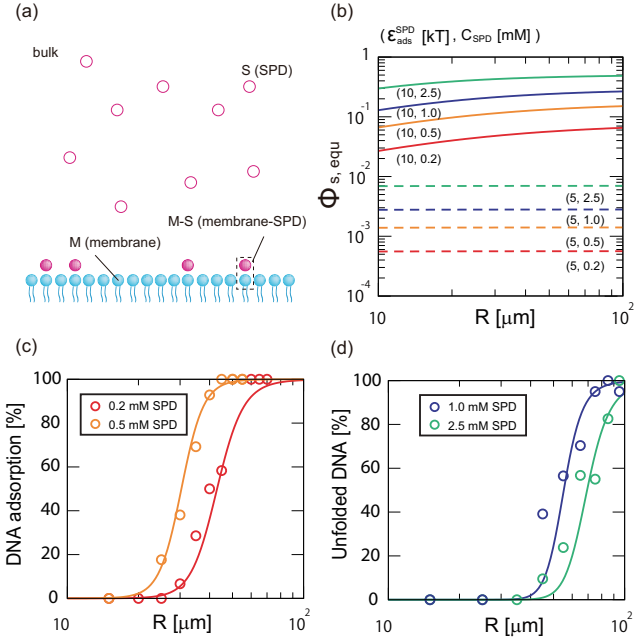}
\caption{\label{1} (a) Schematic illustration of the adsorption of SPD on the membrane. (b) Number of absorbed SPD molecules at thermal equilibrium per lattice cell, $\phi_{\text{s,equ}}$, as a function of $R$ for different adsorption energies of SPD, $\epsilon _\text{ads}^\text{SPD}$, for 5 $k_\text{B}T$, 10 $k_\text{B}T$ and SPD concentrations of 0.2 mM, 0.5 mM, 1.0 mM, and 2.5 mM (see Eq.(\ref{3})). (c) Behavior of coiled DNA with SPD of $\leq $0.5 mM in droplets. Open circles and solid lines represent the experimental data and the theoretical fitting, respectively. (d) Behavior of globular DNA with SPD of $\geq $1.0 mM in droplets. Open circles and solid lines represent the experimental data and the theoretical fitting, respectively \cite{parameters}. }
\end{figure}

We now discuss the mechanism of these trends of DNA behavior in droplets by considering the free energy of DNA. First, we focus on the adsorption transition of coiled DNA onto the membrane with an increase in the droplet size.   
The adsorption transition would be determined by the competition between the gain of free energy due to the adsorption of DNA on the membrane and the loss of free energy due to the change in the translational and conformational entropy of DNA upon moving from three- to two-dimensional space. The change in free energy per molecule is given as 
\begin{eqnarray}
\begin{split}
\Delta F_\text{1}^\text{DNA}= & k_\text{B}T\text{ln}(\frac{r_{2}^{2}}{4\pi R^{2}})- k_\text{B}T\text{ln}(\frac{r_{3}^{3}}{4\pi R^{3}/3})\\
&+\epsilon_\text{conf}-\epsilon _\text{ads}^\text{DNA} \\
                   = & k_\text{B}T\text{ln}(\frac{r_{2}^{2}}{3r_{3}^{3}}R)+\epsilon_\text{conf}-\epsilon _\text{ads}^\text{DNA},
\label{1}
\end{split}
\end{eqnarray}
where $\epsilon_\text{conf}$ is the change in the conformational entropy of DNA, $\epsilon _\text{ads}^\text{DNA}$ is the adsorption energy of DNA, $k_\text{B}$ is the Boltzmann constant, $T$ is temperature, $r_{3}$ is the size of DNA in three-dimensional space, $r_{2}$ is the size of DNA in two-dimensional space, and $R$ is the radius of the droplet (R $\gg$ $r_{3}$, $r_{2}$). 
The first term of Eq.(1) is the change in translational entropy of DNA from three- to two-dimensional space, which is a logarithmically increasing function of $R$ \cite{Kato2009}. 
$\epsilon_\text{ads}^\text{DNA}$ would be an increasing function of $R$,
since $\Delta F_\text{1}^\text{DNA}$ deceases with $R$ and $\epsilon_\text{conf}$ is independent of $R$. 
To consider $\epsilon _\text{ads}^\text{DNA}$ in detail, we studied DNA adsorption on a membrane mediated by SPD.


It is well known that a trivalent positively charged ion such as SPD is markedly absorbed on negatively-charged molecules, and this is referred to as counterion condensation\cite{a,Oosawa}. SPD mediates the attraction between DNA segments that causes DNA to transition into a globular structure  \cite{a,b}. The zeta potential of the DOPC membrane with 10 mM monovalent ions was measured to be -13.4 mV \cite{zeta}. Thus, both DNA and the DOPC membrane should be absorbed by SPD and they have an attractive electrostatic interaction. 
DNA molecules possess a negative bare charge (ca. 600 per Kuhn length with effective ionic dissociation $\sim$ 10 \%), and this charge is reduced by approximately 90\% with SPD in solution due to counterion condensation \cite{Yoshikawa2001}. 
The quantity of SPD needed to neutralize the bare charges of all DNA in a droplet by 90\% is approximately two orders of magnitude less than the entire quantity of SPD. Thus, the quantity of SPD for DNA neutralization is negligible small compared with the total quantity. 
On the other hand, the bare negative charge of the membrane would be partly neutralized by SPD in solution. 

Therefore, we now consider the adsorption equilibrium of SPD on the membrane. For simplicity, we assume that $n$ SPD of the entire quantity, $N$, in the droplet is adsorbed on the membrane with an energy gain of $\epsilon _\text{ads}^\text{SPD}$ (see Fig.4(a)). The total free energy is generally described using the Flory-Huggins model as   
\begin{equation}
\begin{split}
\Delta F_\text{ads}^\text{SPD}=&-n\epsilon_\text{ads}^\text{SPD}+\Omega_\text{s}k_\text{B}T\bigl\{{\phi_{\text{s}}\text{ln}\phi_{\text{s}}+
(1-\phi_{\text{s}})\text{ln}(1-\phi_{\text{s}}) }\bigr\}\\
& + \Omega_\text{v}k_\text{B}T\bigl\{{\phi_{\text{v}}\text{ln}\phi_{\text{v}}+(1-\phi_{\text{v}})\text{ln}(1-\phi_{\text{v}})}\bigr\},
\label{2}
\end{split}
\end{equation}
where $\Omega_\text{s}$ and $\Omega_\text{v}$ are the number of lattice cells on the surface and in the bulk, respectively. With $R$ and the lattice cell length, $a$, corresponding to the molecular length of SPD, $\Omega_\text{s}$ and $\Omega_\text{v}$ can be written as follows: $\Omega_\text{s}=4\pi R^{2}/a^{2}$ and $\Omega_\text{v}=4\pi R^{3}/3a^{3}$. Note that $\phi_{\text{s}}=n/\Omega_\text{s}$ and $\phi_{\text{v}}=(N-n)/\Omega_\text{v}$ are the volume fractions of SPD on the surface and in the bulk, respectively. 
The number of adsorbed SPD molecules at thermal equilibrium per lattice cell, $\phi_{\text{s,equ}} $, is obtained by minimizing $\Delta F_\text{ads}^\text{SPD}$ with respect to $n$ as follows:
\begin{equation}
\begin{split}
 \phi_{\text{s,equ}}(R)= &\frac{1}{2\Omega_\text{s}}\biggl\{N+\Omega_\text{s}+ \Omega_\text{v}e^{-\frac{\epsilon_\text{ads}^\text{SPD} }{k_\text{B}T}} \\
&-\sqrt{(N+\Omega_\text{s}+ \Omega_\text{v}e^{-\frac{\epsilon_\text{ads}^\text{SPD} }{k_\text{B}T}})^2-4\Omega_\text{s}N}\biggr\},
\label{3}
\end{split}
\end{equation} 
where we have assumed the relation $\phi_{\text{v}}\ll 1$. Figure 4(b) shows the dependence of $\phi_{\text{s,equ}}$ on $R$ for different values of $C _\text{SPD}$ used in the experiment and $\epsilon _\text{ads}^\text{SPD}$. Although $\phi_{\text{s,equ}} $ is almost constant under the condition that $\epsilon _\text{ads}^\text{SPD}$ is small (several $k_\text{B}T$), $\phi_{\text{s,equ}} $ is an increasing function of $R$ with $\epsilon _\text{ads}^\text{SPD}$ = 10 $k_\text{B}T$. This increase in free energy can be attributed not only to electrostatic attraction but also to van der Waals attraction.
Since $\epsilon _\text{ads}^\text{DNA}$ increases with the number of SPD adsorbed on the membrane surface, we assume, for simplicity, the linear function $\epsilon _\text{ads}^\text{DNA}=A\phi_{\text{s,equ}}(R)+B$, where $A$ and $B$ are constant. Next, we obtained $\Delta F_\text{1}^\text{DNA}=k_\text{B}T\text{ln}(\frac{r_{2}^{2}}{3r_{3}^{3}}R)-A\phi_{\text{s,equ}}+C$, where $C$ is a constant that is the sum of $-B$ and $\epsilon_\text{conf}$.
The present experimental data show a good fit with this equation (Fig.4(c)) \cite{parameters}. Note that the probability density of the two states is given by a Boltzmann distribution at thermal equilibrium. We used $r_\text{1}=1.5$ $\mu\text{m}$~\cite{r3d1,r3d2} and $r_\text{2}=8.4$ $\mu\text{m}$~\cite{r2d1}. 
Small DNA molecules are expected to adsorb on the membrane surface as T4 DNA does. However, the critical droplet size, above which smaller DNA adsorbs, shifts to a larger size since the adsorption energy decreases as DNA length becomes shorter.


We now turn to the globule-to-coil transition of DNA on the membrane with an increase in the droplet size. The unfolding transition would be determined by the free energy as follows: 
\begin{equation}
\Delta F_\text{2}^\text{DNA}=\Delta F_\text{coil-glo}^\text{DNA}-\epsilon _\text{ads, coil-glo}^\text{DNA}+\epsilon _\text{trans},
\label{4}
\end{equation} 
where $\Delta F_\text{coil-glo}^\text{DNA}$ is the loss of free energy due to the globule-to-coil transition, $\epsilon _\text{ads, coil-glo}^\text{DNA}$ is the difference in adsorption energy of a DNA on the membrane between coiled and globular DNA ($\epsilon _\text{ads, coil-glo}^\text{DNA}=A^{\prime}\phi_{\text{s,equ}}(R)+B^{\prime}$), and $\epsilon _\text{trans}$ is the loss of translational entropy due to the change in DNA conformation. 
Under the condition SPD $\geq $1.0 mM, the energy of the globule state is lower than that of the coil state: $\Delta F_\text{coil-glo}^\text{DNA}$ is an increasing function of $C _\text{SPD}$.
For simplicity, we assume that $\Delta F_\text{coil-glo}^\text{DNA}$ is a linear increasing function of $C_\text{SPD}$ ($\Delta F_\text{coil-glo}^\text{DNA}=D \cdot C_\text{SPD}+E$). With the constant fitting parameters $A^{\prime}$, $D$, and $F$, $\Delta F_\text{2}^\text{DNA}$ is given by $\Delta F_\text{2}^\text{DNA}=-A^{\prime}\phi_{\text{s,equ}}+D \cdot C_\text{SPD}+F$, where F is the sum of $-B^{\prime}$, E, and $\epsilon _\text{trans}$. Our experimental data show a good fit with this equation (Fig.4(d)) \cite{parameters}. 
The adsorption of DNA on lipid membranes has been reported; DNA adsorbed on cationic membranes \cite{Herold, Ange} and on neutral membranes with Mg$^{2+}$ ion \cite{Kato2010,Kato2009}, and adsorption affects the conformation of DNA.

As shown in Fig.3(d), after the unfolding transition, the network structure was formed by the bundling of multiple DNA chains on the membrane. It is well known that a bundle structure with the parallel ordering of DNA chains is produced with a high SPD concentration (SPD of $\geq $1.0 mM) and above an overlapping DNA concentration~\cite{bundle}. Our results indicate that DNA in a confined space tends to exhibit network formation because DNA accumulated on the membrane surface.

\begin{figure}[htbp]
\centering
\includegraphics[scale=0.8]{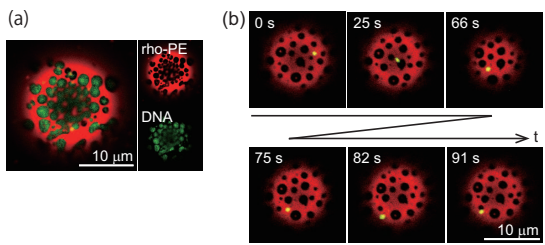}
\caption{\label{1} Selective localization of DNA (green) on phase-separated membrane interfaces, where the membranes were stained with rho-PE (red), which partitions into the disordered phase. (a) Coiled DNA with 0.2 mM SPD was localized in the ordered phase of the membrane. (b) Globular DNA with 1.0 mM SPD was distributed in the disordered phase, and exhibited Brownian motion. The membranes were composed of DPPC and cholesterol (DPPC/cholesterol=1:1) \cite{n1}.}
\end{figure}

Furthermore, we investigated the behavior of DNA molecules in a droplet covered by a heterogeneous membrane surface that mimics biological conditions \cite{JPCL2010, Simons}.  The difference in the fluidity of coexisting phases affects the interaction of membrane-associating molecules \cite{Baumgart, JACS2012, SM2012}. 
Typical fluorescent images of heterogeneous droplet membranes interacting with coiled (0.2 mM SPD) and globular (1.0 mM SPD) DNA are shown in Fig. 5(a) and (b), respectively. Similar to the results with homogeneous membranes, DNA with SPD tended to adhere to the membrane interfaces. We found that there was a clear difference in the localization preference between coiled and globular DNA. Coiled DNA was localized in the ordered phase (Fig. 5(a)), whereas globular DNA selectively interacted with the disordered phase (Fig. 5(b)). The associated globular DNA showed Brownian motion within the disordered region of the phase-separated membrane. 
Since the trend of SPD adsorption on each membrane phase is unclear, the details of free energy are beyond the scope of this paper. However, factors that may contribute to selective adsorption can be considered. The attachment of coiled DNA to the membrane may lead to a loss of translational entropy of lipid molecules, resulting in the localization of DNA on the ordered phase. Conversely, disordered phase-partitioning of globular DNA may be explained in terms of the competition between adsorption energy and the bending elasticity of the membrane. A disordered soft membrane can wrap the associating DNA to overcome the bending energy of the membrane. These considerations have been discussed previously together with the size-dependence of membrane-associating particles \cite{JACS2012}, i.e., larger objects have a softer surface. This is consistent with the present experimental result, in that folded DNA should be treated as a particle that is larger than the segments of coiled DNA. 



\section*{Acknowledgements}
We thank Dr. T. Sakaue and Dr. T. Saitoh for their fruitful discussions, and Ms. A. Hatanaka and Ms. A. Nakade for their technical assistance. This work was supported by MEXT and JSPS KAKENHI (Nos. 23740316, 24115505, 25104510, 26103516, 25103012, 26115709 and 26707020) and by the Kao Foundation for Arts and Sciences, the Kurata Memorial Hitachi Science and Technology Foundation, and a Sunbor Grant from the Suntory Institute for Bioorganic Research.



\begin{thebibliography}{99}
\bibitem{size}
Notably, this results can not be attributed to the time-scale required to achieve an equilibrium configuration, since DNA in small droplets can contact the membrane surface more rapidly by diffusion.
\bibitem{note2}
To construct the phase diagram, we counted the number of droplets with adsorbed DNA (0, 0.2 and 0.5 mM SPD) and unfolded DNA (1.0 and 2.5 mM SPD) to obtain the probabilities as a function of droplet size for each SPD concentration (see also Fig. 4 (c) and (d)). The total number of droplets is N = 94 (0.2 mM SPD), N = 93 (0.5 mM SPD), N = 201 (1.0 mM SPD), and N = 240 (2.5 mM SPD) from two or three repeats of the counting.
\bibitem{n1}
Dipalmitoyl L-$\alpha$ phosphatidylcholine (DPPC), cholesterol (Chol) (Avanti Polar Lipids) and N-(rhodamine red-X)-1,2-dihexadecanoyl-sn-glycero-3-phosphoethanolamine triethylam-monium salt (rho-DHPE) (Invitrogen) were used.
We tested several lipid mixtures (DOPC/DPPC/cholesterol=0/50/50, 50/0/50, 20/60/20) to produce order-disorder lateral phase separation, as well as another fluorescent dye Texas-red DHPE. There was no apparent difference in the phase partitioning of DNA molecules among these membrane materials.
\end{thebibliography}
\end{document}